\documentclass[aps,prd,showpacs, 11pt, notitlepage]{revtex4-1}
\usepackage{graphicx}
\usepackage{booktabs}
\usepackage[utf8]{inputenc}
\usepackage{lmodern}
\usepackage{amsfonts}
\usepackage{amsmath}
\usepackage{amssymb}
\begin{document}

\title{Unitarity and non-relativistic potential energy in a higher-order Lorentz symmetry breaking electromagnetic model}
\author{Eslley Scatena}\email{eslley@feg.unesp.br}
\affiliation{Departamento de F\'isica e Qu\'imica, Universidade Estadual Paulista ``J\'ulio de Mesquita Filho'' - UNESP\\
Av. Ariberto Pereira da Cunha 333, Pedregulho, Guaratinguet\'a, SP - Brazil}
 \author{Rodrigo Turcati}\email{turcati@cbpf.br}
 \affiliation{LAFEX, Centro Brasileiro de Pesquisas F\'isicas - CBPF\\ Rua Dr. Xavier Sigaud 150, Urca, Rio de Janeiro, RJ - Brazil}

\begin{abstract}
The Lorentz-violating model proposed by Myers and Pospelov suffers from a higher-derivative pathology due to a dimension-5 operator. In particular, its electromagnetic sector exhibits an spectrum which contains, in addition to an expected massless photon, ghost contributions that could (in principle) spoil the unitarity of the model. We find that unitarity at tree-level can be assured for pure spacelike, timelike and lightlike background four-vectors (the last two under restrictions upon the allowed momenta). We then analyze the non-relativistic interparticle potential energy behavior for different background four-vectors and compare to the usual Coulomb potential.
\end{abstract}

\maketitle

\section{Introduction}

The search for new physical effects that may emerge at high-energy process or could arise at a quantum gravity level has lead some physicists to propose modifications to the Standard Model of particles and General Relativity. The current view is that our perturbative approach that works so well in explaining the accessible energy scale nowadays is just an \emph{effective theory}, meaning that it is the lower limit of an underlying unified theory. However, since we are still not able (and will not be in an foreseeable future) to experimentally probe this Planck scale ($\approx 10^{19} GeV$), we are actually searching for small deviations from the standard theories as such those suppressed by the Planck mass, for instance \cite{AmelinoCamelia:1997gz, Ellis:1999rz}.

If we expect to find this type of deviation that comes from an unified theory (\emph{e.g.}, string theory, loop quantum gravity, non-commutative field theory \cite{Kostelecky:1988zi,AmelinoCamelia:1999pm}), one should asks ``which is the most fundamental aspect of our actual theories that may not hold in an unified scheme and how it would manifest itself as we get closer to the limit of validity of our effective theory?'' In the context of quantum field theories, the Lorentz invariance is one of the greatest foundations that have been put into scrutiny recently (ironically, the same invariance that took us from classical to relativistic physics in the past century) \cite{lrr-2005-5}.

In this vein, Kosteleck\'y and collaborators has carried out an systematic program in order to classify and quantitatively describe Lorentz and CPT symmetry violations, parametrized by a set of coefficients determined by experiments, namely, the Standard Model Extension (SME)\cite{Colladay:1998fq}. However, since the various sectors of the usual Standard Model ($SU(3)\times SU(2)\times U(1)$) exhibit a plethora of those coefficients, it would be wise to have a guide principle to carefully study those violations. An interesting proposal made by Myers and Pospelov (based in six consistent criteria) considers ultraviolet modifications in the photon's dispersion relation induced by a dimension-5 operator that, besides an external background field that violates Lorentz invariance, also presents higher derivatives \cite{Myers:2003fd}. 

As it is expected, the presence of this higher-derivative operator introduce ghost states that, in principle, could jeopardize the unitarity of the model (since the energy wouldn't be bounded from below). The problems of non-unitarity concerning higher-derivative theories has been extensively studied in the context of Lee and Wick theories, and many solutions for this issue has been proposed along the last decades \cite{Lee:1969fy,Lee:1971ix,Nakanishi:1972pt, reyes2,Shalaby:2009re}. Although it was recently shown, for the Myers-Pospelov model, that in electron-positron and Compton scattering there is no contribution of the ghost states \cite{Reyes:2013nca,Maniatis:2014xja}, we present here a alternative approach to the subject of tree-level unitarity.

Based on an method pioneered by Veltman that has been extensively used even in theories violating Lorentz symmetry (\emph{e.g.} Maxwell-Proca-Chern-Simons or Carroll-Field-Jackiw theories), we analyze the poles of the saturated propagator of the model in order to check the unitarity at tree-level \cite{Veltman:1994wz,PhysRevD.67.085021}. We also take advantage of the previous calculated saturated propagator to find deviations from the usual potential energy between two charges, since a modified dispersion relation can be translated into modifications to the Coulomb's law \cite{0034-4885-68-1-R02,Goldhaber:2008xy}.

In order to achieve those results, in Section 2 we show how we obtain the Feynman propagator for the electromagnetic sector of the Myers-Pospelov model and find the conditions for the model to be unitary. In Section 3 we proceed to extract the non-relativistic limit of the potential energy between two static point charges, and calculate this potential for different background vectors. We conclude in Section 4 with some considerations about the implications of our results and further analysis.

In our conventions $\hbar = c =1$, and  the metric signature is (+ - - -). 

\section{The Feynman propagator of the model and unitarity}

The free electromagnetic sector of the Myers-Pospelov model is defined by the following lagrangian
\begin{equation}\small
\mathcal{L}=-\frac{1}{4}F^{\mu\nu}F_{\mu\nu}+\frac{g}{2}n^{\mu}F_{\mu\nu}(n\cdot\partial)n_{\alpha}\tilde{F}^{\alpha\nu}-\frac{1}{2\lambda}\left(\partial^{\mu}A_{\mu}\right)^2,\label{lgamma}
\end{equation}
where $F_{\mu\nu}=\partial_{\mu}A_{\nu}-\partial_{\nu}A_{\mu}$ is the usual field strength of the electromagnetic field and $n^{\alpha}$ is the Lorentz violating four-vector defining a preferred reference frame. Moreover, $g=\frac{\xi}{M_{p}}$, with $\xi$ being a dimensionless parameter suppressed by the Planck mass $M_{p}$, and $\lambda$ is a gauge parameter.

Notwithstanding, to analyze the unitarity of the model and find the non-relativistic interparticle potential energy, we need to find the Feynman propagator that intermediates this interaction. The above lagrangian can be rewritten as

\begin{equation*}
\mathcal{L}=\frac{1}{2}A_{\nu}\Delta^{\nu\sigma}A_{\sigma}, 
\end{equation*}
with
\begin{equation*}
\Delta^{\nu\sigma}=\left[\Box\eta^{\nu\sigma}-\partial^{\nu}\partial^{\sigma}\left(1-\frac{1}{\lambda}\right)-2gn_{\alpha}\epsilon^{\alpha\nu\rho\sigma}\left(n\cdot\partial\right)^2\partial_{\rho}\right],
\end{equation*}
where $\Delta^{\nu\sigma}$ is the wave operator associated with the lagrangian. A formal inversion of the operator $\Delta$ in the momentum space will give us the correct form of the propagator
\begin{eqnarray} 
(\Delta^{-1})^{\mu\nu}&=& \frac{1}{D(k)}\Big\{-k^{2}\eta^{\mu\nu}-4g^{2}(n\cdot k)^{4}n^{2}\omega^{\mu\nu}+\nonumber\\&-&4g^{2}(n\cdot k)^{4}\left[n^{\mu}n^{\nu}-\frac{(n\cdot k)}{k^{2}}(k^{\mu}n^{\nu}+n^{\mu}k^{\nu})\right]\nonumber\\
&+&2gi(n\cdot k)^{2}\varepsilon^{\mu\nu\alpha\beta}n_{\alpha}k_{\beta}\Big\}, \label{propagator}
\end{eqnarray}
with $D(k)=k^{4}-4g^{2}(n\cdot k)^{4}[(n\cdot k)^{2}-n^{2}k^{2}]$.
If we saturate the propagator in (\ref{propagator}) with conserved currents, i.e., $SP\equiv J^{\mu}(k)\Delta^{-1}_{\mu\nu}J^{\nu}(k)$, we are left with
\begin{equation}
J^{\mu}\Delta^{-1}_{\mu\nu}J^{\nu}= \frac{-k^{2}J^2-4g^{2}(n\cdot k)^{4}(n\cdot J)^2}{k^{4}-4g^{2}(n\cdot k)^{4}[(n\cdot k)^{2}-n^{2}k^{2}]}.\label{saturated}
\end{equation}

In the above expression, $J^{\mu}(k)\equiv\int d^4x e^{-ikx} J^{\mu}(x)$ is the conserved four-current such that $k^{\mu}J_{\mu}(k)=0$. We can assure the tree-level unitarity of our model if the residues of the saturated propagator ($SP$) calculated in its simple poles are greater than zero for propagating modes \cite{Veltman:1994wz}.
Since we can have three distinct situations for the Lorentz-violating background four-vector, we proceed our analysis with each one of those possibilities separately. 

\subsection{Spacelike Lorentz-violating background four-vector}

If we choose a representation such that $k_{\mu}=(k_0,0,0,k_3)$, thus the current conservation implies $J_{\mu}=(J_0,J_1,J_2,\frac{J_0k_0}{k_3})$. Taking these results into account and choosing a pure spacelike background four-vector such as $n_{\mu}=(0,0,0,1)$, the saturated propagator (\ref{saturated}) assumes, with no loss of generality, the following form
\begin{equation*}
SP= \frac{-k^{2}J^2-4g^{2}k_3^{4}J_3^2}{(k^2-m^2_{+})(k^2-m^2_{-})},
\end{equation*}
with the poles $m^2_{\pm}=k_0^2-k^2_3=2g^2k_3^4\pm2gk_3^3\sqrt{1+g^2k_3^2}$. 
Therefore, the residue of the saturated propagator in each pole is
\begin{eqnarray*}
\text{Res}[SP]|_{k^2=m^2_{\pm}}=
\left[\frac{\pm 1}{2\sqrt{1+\frac{1}{g^2k_3^2}}}+\frac{1}{2}\right]\left(J_1^2+J_2^2\right)>0.
\end{eqnarray*}

As can be seen from the previous result, the residue is always positive for any real value of $k_3$, thus ensuring the tree-level unitarity of the model in the spacelike case.

\subsection{Timelike Lorentz-violating background four-vector}

The timelike situation is, \emph{mutatis mutandis}, similar to the previous case. We use a pure timelike background four-vector $n_{\mu}=(1,0,0,0)$ and the saturated propagator (\ref{saturated}) assumes the form

\begin{equation*}
SP= \frac{-k^{2}J^2-4g^{2}k_0^{4}J_0^2}{(k^2-m^2_{+})(k^2-m^2_{-})}.
\end{equation*}
Here, the poles $m^2_{\pm}$ are solutions of $D(k)$ for the timelike $n_{\mu}$, and can be written as
\begin{equation*}
m^2_{\pm}=\frac{k_3^2}{1\pm 2gk_3}-k_3^2.
\end{equation*}
The residue of the saturated propagator in those poles yields
\begin{eqnarray*}
\text{Res}[SP]|_{k^2=m^2_{\pm}}=
%
\frac{J_1^2+J_2^2}{2\left(1\pm2gk_3\right)\left(1-4g^2k_3^2\right)}.
\end{eqnarray*}

Contrary to the spacelike case, we have found that the sign of $\text{Res}[SP]|_{k^2=m^2_{\pm}}$ depends upon the the sign of $(1\pm2gk_3)$, which could introduce a ghost in the spectrum and spoil the unitarity of the model in this particular case. However, this contribution can be discarded since its energy ($\sim 1/2g$) lies beyond the region of validity of the effective theory (we expect $1/2g$ to be comparable to the Planck scale), and we can restore  $\text{Res}[SP]|_{k^2=m^2_{\pm}}>0$ for $|k_3|<\frac{1}{2g}$. Therefore one must not be afraid, for this region is not haunted by ghosts.

\subsection{Lightlike Lorentz-violating background four-vector}\label{sectionll}

From the previous analysis of different types of backgrounds, we can see that the general form of the residues of the saturated propagator in each pole $m^2_{i}$ cancels out the $J_0$ and $J_3$ contributions (due to current conservation), giving

\begin{equation*}
\text{Res}[SP]|_{k^2=m_{i}^2}=\frac{m_{i}^2\left(J_1^2+J^2_2\right)}{\prod_{j=1}^{n}\left(m_{i}^2-m_{j}^2\right)}, \quad j\neq i,
\end{equation*}
where the $m^2_{j}$'s are the other $n$ roots of $D(k)$.

\begin{table}\centering
\begin{tabular}{lll}\hline
 Pole & $Res[SP]|_{m^2_{(i)}}/(J_1^2+J_2^2)$ & Conditions for $Res[SP]|_{m^2_{(i)}}>0$\\
\hline
$m^2_{(1)}$ & $\frac{1}{k_3}$& $k_3>0$\\

$m^2_{(2)+}$ & $\frac{2 g}{1 - 16 g k_3 - \sqrt{1 - 16 g k_3}}$ &  $k_3<0$\\

$m^2_{(2)-}$ & $-\frac{1}{8k_3}+ \frac{1}{8k_3\sqrt{1 - 16 g k_3}}$ & $k_3<0$, or $0 <k_3< \frac{1}{16g}$\\
$m^2_{(3)+}$ & $-\frac{1}{8k_3}-\frac{1}{8k_3\sqrt{1 + 16 g k_3}}$ & $k_3 > 0$ or $-\frac{1}{16 g} < k_3 < 0$\\

$m^2_{(3)-}$ & $-\frac{1}{8k_3}+\frac{1}{8k_3\sqrt{1 + 16 g k_3}}$ & $-\frac{1}{16 g} < k_3 < 0$\\
\hline

\end{tabular}
\caption{Residues for different poles of the saturated propagator in the lightlike case ($n^2=0$), and conditions for $Res[SP]|_{m^2_{(i)}}>0$.}
\label{tabela1}
\end{table}

For the lightlike case, with $n_{\mu}=(1,0,0,1)$ as a preferred background, the poles in which we are interested in are the solutions of $D(k)=k^4-4g^2(k_0-k_3)^6=0$, 

\begin{eqnarray*}
&m^2_{(1)}&=0;\\
&m^2_{(2)\pm}&=\left[\frac{-1+ 4gk_3\pm\sqrt{1-16gk_3}}{4g}\right]^2-k_3^2;\\
&m^2_{(3)\pm}&=\left[\frac{1+ 4gk_3\pm\sqrt{1+16gk_3}}{4g}\right]^2 -k_3^2.
\end{eqnarray*}

Just like in the timelike situation, there are some troublesome poles that could entail negative norm states in the model, as can be directly seen in Table \ref{tabela1}. Therefore, for $|k_3|<\frac{1}{16g}$, there are attainable real $m_{i}$ excitations such that $Res[SP]|_{m^2_{(i)}}>0$, preserving the unitarity as well.

\section{Non-relativistic potential energy}
One important question to pose, since we are searching for ways to probe the effects of a Lorentz-symmetry violation, is whether classical effects can be sensible to this violation or not. Although those effects are suppressed by the Planck scale, it is worth to see what kind of deviation from the Coulomb potential this violation would introduce, how sensitive this correction is and if any other effect may emerge at this non-relativistic level.

With the explicit form of the propagator given in (\ref{propagator}), we are able to compute the non-relativistic potential energy between two static point charges separated by a distance $\mathbf{r}=\mathbf{x}_1-\mathbf{x}_2$.

The Yukawa-like potential in which we are interested in can be found using the method based on the path integral formalism. Since the generating functional related to the connected Feynman diagrams, $W(J)$, is related to the generating functional for the free theory, $Z(J)$, by $ Z(J)= e^{iW(J)}$, we find \cite{Zee:2003mt}

\begin{eqnarray}
W(J)= -\frac{1}{2}\int \frac{d^4k}{(2\pi)^4}J^{\mu}(k)^*\Delta^{-1}_{\mu\nu}(k)J^{\nu}(k).\label{potential1}
\end{eqnarray}

For a charge distribution $J^{\mu}(x)= \eta^{0\mu} [q_1\delta^{3}({\bf {x}} - {\bf{x_1}}) + q_2\delta^{3}({\bf {x}} - {\bf{x_2}})]$, we have two static $q_{i}$ ($i=1,2$) point charges separated by $\bf{r}$. 

Substituting the charge distribution into (\ref{potential}) and bearing in mind that in the path integral formalism we have $Z=e^{iW(J)}=\left\langle0|e^{-iHt}|0\right\rangle =e^{-iEt}$ (where $E$ is the interparticle energy that we want to find), we have $iW=iEt$. After integrate the 0th components, we obtain
\begin{equation}
E=\int\frac{d^3\mathbf{k}}{(2\pi)^3}\frac{q_1q_2\left[|\mathbf{k}|^2+4g^2(\mathbf{n}\cdot\mathbf{k})^4n_0^2\right]e^{i\mathbf{k}\cdot\mathbf{r}}}{|\mathbf{k}|^4-4g^2(\mathbf{n}\cdot\mathbf{k})^4\left[(\mathbf{n}\cdot\mathbf{k})^2+n^2|\mathbf{k}|^2\right]}\label{potential}.
\end{equation}

\subsection{Timelike potential energy}
If we proceed our analysis for the timelike case (with $n_{0}=1, \mathbf{n}=0$), we find that it gives us no additional information, since the potential energy in (\ref{potential}) reduces to the Coulomb one, \emph{i. e.},
\begin{eqnarray*}
E_{tl}=q_1q_2\int\frac{d^3\mathbf{k}}{(2\pi)^3}\frac{e^{i\mathbf{k}\cdot\mathbf{r}}}{|\mathbf{k}|^2}=\frac{q_1q_2}{4\pi}\frac{1}{r}.
\end{eqnarray*}
This result is somewhat expected, taking into account that the pure timelike case does not introduce any anisotropies in space. Therefore, we cannot observe any modifications introduced by the Lorentz-violating background vector for this non-relativistic approximation.

\subsection{Spacelike potential energy}
Using a similar procedure as before, we define a pure spacelike four-vector $n_{\mu}=(0,\mathbf{n})$ and consider the following relations (for $n\|r$):
\begin{equation}
\mathbf{n}\cdot\mathbf{k}=nk\cos{\theta}, \quad \mathbf{k}\cdot\mathbf{r}=kr\cos{\theta}, \quad\text{and}\quad \mathbf{n}\cdot\mathbf{r}=nr.
\end{equation}
With those assumptions we are left only with the first term of (\ref{potential}) (since $n_0=0$), which can be expressed in spherical coordinates as
\begin{equation}
E_{sl}=\frac{q_1q_2}{(2\pi)^2}\int_0^{\infty}\int_0^{\pi}\frac{e^{ikr\cos{\theta}}\sin{\theta}}{1+4g^2k^2n^6\cos^4{\theta}\sin^2{\theta}}d\theta dk.
\end{equation}
Performing an expansion in $k^2g^2 (\ll 1)$ and taking $|\mathbf{n}|=n=1$, we have, at second order in $g$
\begin{equation*}
\frac{1}{1+4g^2k^2\cos^4{\theta}\sin^2{\theta}}\approx(1-4g^2k^2\cos^4{\theta}\sin^2{\theta}),
\end{equation*}
and we are left with
\begin{eqnarray}
E_{sl}&=&\frac{q_1q_2}{(2\pi)^2}\int_0^{\infty}\int_0^{\pi} e^{ikr\cos{\theta}}\sin{\theta}\left[1+\right. \nonumber\\
&-&\left.4g^2k^2\cos^4{\theta}\sin^2{\theta}\right]d\theta dk.
\end{eqnarray}
In this way, the correction to the ordinary Coulomb potential energy $E_{C}(=\frac{q_1q_2}{4\pi r})$ will be
\begin{eqnarray}
E_{sl}&=&E_{C}-E^{g}_{sl}, \text{ where} \nonumber\\
E^{g}_{sl}&=&\frac{4g^2q_1q_2}{(2\pi)^2}\int_0^{\infty}\int_0^{\pi}e^{ikr\cos{\theta}}k^2\cos^4{\theta}\sin^3{\theta}d\theta dk.\label{ug}
\end{eqnarray}
As one can see, this last term is highly divergent, so we have to introduce a cutoff in order to obtain a meaningful result to analyze. In this approximation an appropriate choice would be $\Lambda=1/16g$, since, as it was shown in Section \ref{sectionll}, we can preserve the unitarity of the model in the timelike and lightlike situations provided $|\mathbf{k}|<1/(16g)$ (a discussion about the effectiveness of the quantum model at this scale can be found in \cite{Reyes:2008rj}). With this cutoff, keeping only terms up to $g^2$, the equation (\ref{ug}) reduces to
\begin{eqnarray}
E_{sl}^{g}&=&-\frac{16g^2q_1q_2}{(2\pi)^2r^3}\sin{\left(\frac{r}{16 g}\right)}
\end{eqnarray}
and the potential energy takes the form
\begin{eqnarray}
\nonumber E_{sl}&=&\frac{q_1q_2}{4\pi r}\left\{1+\frac{16g^2}{\pi r^2}\sin{\left(\frac{r}{16 g}\right)}\right\}.
\end{eqnarray}
We can see that deviations from Coulomb's law are suppressed by a factor of $g^2$ and smoothly reduces to it for $\lim_{g\rightarrow 0} E_{sl}^{g} =0$. Using the existent bounds on $\xi$ ($<10^{-15}$) and remembering that $g=\xi/M_{P}$, it can be found $g\sim 10^{-41}m$. Even less stringent limits ($\approx 1$) sets $g$ to $\sim 10^{-26}m$ \cite{1475-7516-2008-08-027,PhysRevLett.100.021102,PhysRevD.82.024013}. Therefore, as expected, we shouldn't observe any departure from the Maxwellian potential in the range of validity of this semiclassical approximation, since deviations of the ordinary potential $1/r$ only takes place for distances comparable to the Planck length ($l_{P}\sim 10^{-35}m$).

\begin{figure}
  \begin{center}
    \includegraphics[scale=0.5]{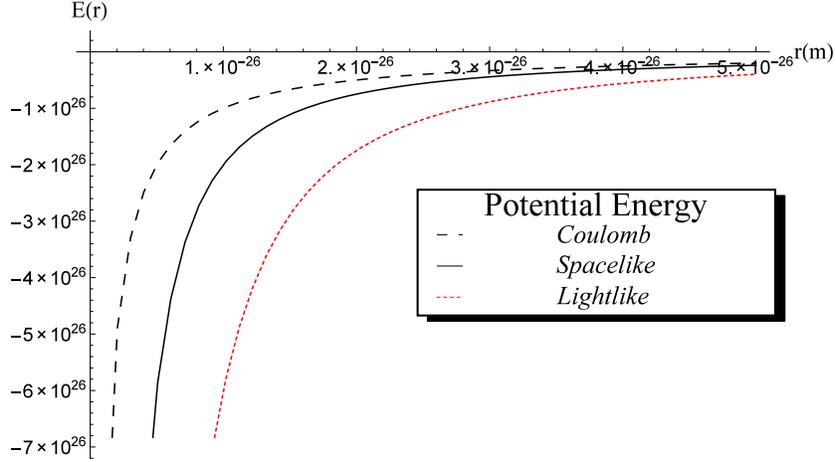}
    \caption{Plot of the potential energy between two unitary charges $q_1=1=-q_2$ for the Coulomb case $E_{C}$ (solid line), for the spacelike case $E_{sl}$(dashed line), and for the lightlike case $E_{ll}$ (dotted line), with $g=1\times10^{-26}m$.}
    \label{figure1}
  \end{center}
\end{figure}

\subsection{Lightlike potential energy}
In the case for a lightlike background four-vector we have the additional contribution of the second term in (\ref{potential}), since now $|n_0|^2=|\mathbf{n}|^2=1$, and the potential energy takes the form 

\begin{equation*}
E_{ll}=\frac{q_1q_2}{(2\pi)^2}\int_0^{\infty}\int_0^{\pi}\frac{1+4g^2k^2\cos^4{\theta}}{1-4g^2k^2\cos^6{\theta}}e^{i\mathbf{kr\cos{\theta}}}\sin{\theta}d\theta dk.
\end{equation*}
Carrying a similar expansion and keeping terms of order $g^2$, we can write

\begin{eqnarray}
E_{ll}&=&E_{C}+E_{ll}, \text{ where}\\
E_{ll}^{g}&=&\frac{4g^2q_{1}q_{2}}{(2\pi)^2}\int_0^{\Lambda}\int_0^{\pi}e^{i\mathbf{kr\cos{\theta}}}k^2(1+\cos^2{\theta})\cos^4{\theta}\sin{\theta}d\theta dk.\nonumber
\end{eqnarray}
Integrating $E_{ll}^{g}$ enables us to write 

\begin{eqnarray}
E_{ll}=\frac{q_{1}q_{2}}{4\pi r}\left\{1-\frac{g}{\pi r}\left[\cos{\left(\frac{r}{16g}\right)}-\frac{96g}{r} \sin{\left(\frac{r}{16g}\right)}\right]\right\}.\nonumber
\end{eqnarray}
For values of $r$ comparable to $g$ we approximate the cutoff and the solution above is no longer valid. As we can see, it oscillates since now we have significant contributions that arise from $\cos{\left(r/2g\right)}$ and $\sin{\left(r/2g\right)}$. However, in its region of validity (\emph{i.e.}, the Compton wavelength of the electron, $\sim 10^{-12}m$, although the Coulomb's law is verified up to $\sim 10^{-17}m$ \cite{0034-4885-68-1-R02}), the potential is essentially Coulombian (as it should for $g\rightarrow 0$ as well). It is also important to stress here that the values adopted for $g$ (or $\xi$) doesn't necessarily have to be the same for the different backgrounds, as observed in \cite{PhysRevD.82.024013}, but even for the less stringent limits we found no deviations.

\section{Discussion}
In this paper we have studied the unitarity of the electromagnetic sector of a Lorentz-violating model with a dimension-5 operator proposed by Myers and Pospelov. Since this operator introduces higher-derivative terms, it is possible that negative norm states emerge upon the choice of the background four-vector $n_{\mu}$. Analyzing under which conditions the residue of the saturated propagator is positive, we found a class of possible results that are unitary at tree-level. For the \emph{spacelike} case the conditions for unitarity are completely fulfilled. Moreover, we found that for the \emph{lightlike} and \emph{timelike} situations, a cutoff can be implemented and the unitarity can be assured, confirming the results obtained by Reyes for the electron-positron scattering  \cite{Reyes:2013nca}.
Taking advantage of the saturated propagator obtained before, we also found the non-relativistic potential energy from the interaction between two charges. 

In the process for obtaining those results we have introduced a cutoff $\Lambda=1/16g$, based on the constraint that guarantees the unitarity in both \emph{timelike} and \emph{lightlike} situations. Since we can always perform a boost such that $n_{\mu}=(n_0,0,0,0)$ acquires a spacial component, we must restrict the possible concordant frames as well  \cite{Reyes:2010pv}, which implies a cutoff for other anisotropic cases. It is also important to notice that those results, even though obtained for particular $n_{\mu}$, signals that we shouldn't expect any modifications in the general potential energy given by (\ref{potential}). Furthermore, in its complete form we should expect a dependence upon the constant angle $\alpha$ between $\mathbf{n}$ and $\mathbf{r}$ ($\mathbf{n\cdot r}=nr\cos{\alpha}$), making explicit the induced space anisotropy.

It is clear that significant contributions will only manifest themselves in higher energy processes or some quantum effect, so that we can really grasp the modifications introduced by this Lorentz symmetry violation. From a theoretical point of view, we should be able to understand what are the mechanisms that induce such a violation and what is the best framework to study them. In this vein, the emerging scenario of VSR (\emph{Very Special Relativity}) seems to accommodate very well those kind of symmetry breaking \cite{Cohen:2006ky,Alfaro:2013uva}. Recently it was also shown that one can obtain the Myers-Pospelov model by introducing the fermion sector of a Lorentz-symmetry violating master QED and radiatively inducing a master effective action \cite{Mariz:2011ed,Anacleto:2014aha}.

\section*{Acknowledgments}
We would like to thank professors A. Accioly, J. M. Hoff da Silva and M. Hott for useful comments and discussion.
E.S. thanks the Brazilian agency CAPES/PNPD and the Dept. of Physics and Chemistry (DFQ-FEG) for the full support. The work of R.T. is supported by the Science Without Borders fellowship, from the Brazilian agency CNPq.

\bibliography{Unitarity}

\end{document}